\documentclass[twocolumn,showpacs,preprintnumbers,amsmath,amssymb]{revtex4}
\usepackage{graphicx}
\usepackage{dcolumn}
\usepackage{bm}

\newcommand{\bef}{\begin{figure}}
\newcommand{\eef}{\end{figure}}
\newcommand{\nn}{\nonumber}
\newcommand{\be}{\begin{equation}}
\newcommand{\ee}{\end{equation}}
\newcommand{\bea}{\begin{eqnarray}}
\newcommand{\eea}{\end{eqnarray}}

\newcommand{\la}{\langle}
\newcommand{\ra}{\rangle}
\begin{document}

\title{Effect of anisotropy on HBT radii using leptonpair interferometry}



\author{Payal Mohanty$^{1,2}$}
\email{payal.mohanty@gmail.com}
\author{Mahatsab Mandal$^1$} 
\email{mahatsab.mandal@saha.ac.in}
\author{Pradip Roy$^1$}
\email{pradipk.roy@saha.ac.in}

\medskip

\affiliation{$^{1}$High Energy Nuclear and Particle Physics Division, Saha Institute of Nuclear Physics,
  1/AF Bidhannagar, Kolkata-700 064, INDIA, \\ 
$^2$Institute of Physics and Applied Physics, Yonsei University, Seoul 120-749, Korea}

\date{\today}

\begin{abstract}
The effect of initial state momentum-space anisotropy on invariant mass dependence  of HBT radii 
extracted from the leptonpair interferometry is presented here. We have studied the Bose-Einstein 
Correlation Function (BECF) for two identical virtual photons decaying 
to leptonpairs  at most central collision of LHC energy having fixed transverse momentum of one of the 
virtual photons ($k_{1T}$= 2 GeV). The {\em free streaming interpolating} model with fixed initial 
condition has been used for the evolution in anisotropic Quark Gluon Plasma (aQGP) and the relativistic 
(1+2)d hydrodynamics model with cylindrical symmetry and  longitudinal boost invariance  has been 
used for both isotropic Quark Gluon Plasma (iQGP) and hadronic phases. We found a significant change in the 
spatial and temporal dimension of the evolving system in presence of initial state momentum-space anisotropy.
\end{abstract}

\pacs{25.75.+r,25.75.-q,12.38.Mh}
\maketitle
\section{Introduction}
Two-particle intensity interferometry, commonly known as HBT interferometry is one of the efficient 
ways to know the space-time description of particle emission zone created in high energy nucleus-nucleus 
collisions~\cite{pratt,uaw,hb3}. This method was formulated and exploited by Hanbury, Brown and Twiss  
to measure the angular diameter of astronomical objects~\cite{HBT}. It was first introduced in heavy ion collision (HIC)
through  pion interferometry which provided valuable information about the space-time description of 
the system  at the freeze out surface only~\cite{uaw}. It has been argued that in contrast to hadrons, two-particle intensity 
interferometry of photons and dileptons ~\cite{peressounko,fmann,AlamHBT,BMSPRL,DKS_HBT,Rupa_HBT,payal_HBT,photon_ans} 
which are produced throughout the space-time evolution of the reaction zone and which suffer almost no interactions with 
the surrounding medium can provide information on the the history of the evolution of the hot matter created in HIC. 
However photons appear to be a more restrictive probe since they are characterized only by their transverse momentum ($p_T$) 
whereas the dileptons have two kinematic variables, $p_T$ and invariant mass(M) to play with. A soft photon (low $p_T$ ) 
in one frame of reference can be hard (high $p_T$ ) in another frame, whereas  $p_T$ integrated invariant mass 
distribution of dileptons is independent of any frame. In addition to it,  $p_T$ spectra is affected
by the flow, however,  $p_T$ integrated M spectra remain unaltered by the flow in the
system~\cite{payal_flow}. Also in the M spectra of dileptons, above $\phi$ peak, dileptons from QGP dominates
over its hadronic counterpart~\cite{payal_v2}. All these suggests that a judicious choice of $p_T$ and M
windows will be very useful to characterize the QGP and hadronic phase separately.  
Moreover owing to rapid longitudinal expansion at the onset of QGP phase compared to the partonic interaction rate, 
the anisotropy arises in $p_T-p_L$ plane with $\la p_L^2 \ra \ll \la p_T^2 \ra $ in the local rest frame. 
With time,  such asymmetry dies out with secondary partonic interactions. After which the system is considered 
to be isotropic and thermal at proper time $\tau_{iso}$ and beyond $\tau\geq\tau_{iso}$  the system can be treated 
hydrodynamically. 
To include such momentum anisotropy in pre-equilibrium stage of QGP, a simple phenomenological model is adopted from  
refs.~\cite{MMprl100,mauricio,schenke}. 
We assumed two time scales here; (i) the initial QGP formation time, $\tau_i$, 
and (ii) the isotropization time, $\tau_{iso}$, when the isotropy in momentum space is
achieved and they should fulfill the criteria  $\tau_i \leq \tau_{iso}$. In absence of 
anisotropy, we have $\tau_i = \tau_{iso}$. 

In this letter, we present intensity 
interferometry with leptonpairs at most central LHC initial conditions 
at $\sqrt{s}$=2.76 TeV including momentum space anisotropy in pre-equilibrium QGP phase and
attempt to study the effect of the anisotropy on the mass dependence of HBT radii extracted from the Bose Einstein 
Correlation Function, $C_2$ for two identical virtual photons which later decay into leptonpairs. We have discussed the 
definition and formalism of leptonpair interferometry in Sec.~\ref{sec_defn}, the results are given in Sec.~\ref{sec_results} 
and finally we have summarized in Sec.~\ref{sec_summ}.

\section{Definition and Formalism}
\label{sec_defn}
Leptonpair interferometry is based on 
computing the Bose-Einstein correlation (BEC) function for two 
identical virtual photons which later decay in leptonpairs and can be defined as~\cite{pratt,payal_HBT},
\begin{widetext}
\begin{equation}
C_{2}(\vec{k_{1}},\vec{k_{2}})=\frac{P_{2}(\vec{k_{1}},\vec{k_{2}})}
{P_{1}(\vec{k_{1}}) P_{1}(\vec{k_{2}})} 
\label{eq_c2}
\end{equation}

where
\begin{equation}
P_{1}(\vec{k_i}) = \int d^{4}x~\int dM_i^2\,\omega(x,k_i)
\label{eq_P1}
\end{equation}
and

\begin{equation}
P_{2}(\vec{k_{1}},\vec{k_{2}})= 
P_{1}(\vec{k_{1}})P_{1}(\vec{k_{2}})+
\frac{1}{3}\int d^{4}x_{1} d^{4}x_{2}~ dM_1^2 ~dM_2^2  ~\omega (x_{1},K) 
\omega (x_{2},K)~\cos(\Delta x^{\mu} \Delta k_{\mu})
\label{eq_P2}
\end{equation}
\end{widetext}
where $\vec{k_i}=(k_{iT}\cos \psi_i, k_{iT}\sin \psi_i,k_{iT}\sinh y_i)$ 
is the three momentum of the two identical virtual photons with $i=1,2$,  
$K=(k_1+k_2)/2$ is the average transverse momentum, 
$\Delta k_\mu=k_{1\mu}-k_{2\mu}=q_\mu$,
$x_i$ and $k_i$ are the four co-ordinates for position and momentum 
variables respectively and
$\psi_i$'s are the angles made by $k_{iT}$ with the x-axis of each virtual photon. 
 $\omega(x,k)=dR/dM^2d^2k_Tdy$ is the source function 
related to the thermal emission rate of virtual photons  per unit four volume. 
 The possiblity of dilution of the signal due to random pairs 
will not affect the HBT radii (disscussed in ref~\cite{payal_HBT}), thus we have neglected the 
leptonpairs with different invariant masses.
By ignoring the leptonpair with different inavarant mass, i.e, $M_1=M_2=M$, we can re write 
$C_2$ as:
\begin{widetext}
\begin{equation}
 C_2(\vec{k_{1}},\vec{k_{2}})=1+\frac
 {\left[\int d^{4}x ~dM^2  ~\omega (x,K) \cos(\Delta\alpha)\right]^2 +
 \left[\int d^{4}x ~dM^2  ~\omega (x,K) \sin(\Delta\alpha)\right]^2}
 {P_{1}(\vec{k_{1}}) P_{1}(\vec{k_{2}})} 
 \label{eq_c2_new}
\end{equation}
\end{widetext}
where $\Delta\alpha=\alpha_1-\alpha_2$ and $\alpha_i=\tau M_{iT}\cosh(y_i-\eta)-rk_{iT}\cos(\theta-\psi_i)$, 
$M_{iT}=\sqrt{M^2+k_{iT}^2}$
The inclusion of the spin of the  virtual photon  
will reduce the value of $C_2-1$ by 1/3. 
The correlation functions can be evaluated  by using 
Eqs.~\ref{eq_c2}, \ref{eq_P1}, ~\ref{eq_P2} and ~\ref{eq_c2_new}
for different average mass windows,
$\langle M\rangle$. 
We follow Ref.~\cite{MMprl100,mauricio} for the dilepton production in aQGP.
Beyond $\tau \geq \tau_{iso}$, the leading order process through which lepton 
pairs are produced in QGP is $q\bar{q}\rightarrow l^+l^-$~\cite{qqpair}. 
For  low $M$, dilepton production from the hadronic phase the decays of the light vector mesons 
$\rho, \omega$ and $\phi$ have been considered including the continuum~\cite{emprobe,shu}.  
Since the continuum part of the vector meson spectral functions are included,   
the processes like four pions annihilation~\cite{4pi} are excluded to avoid 
double counting. 

\begin{table}[!hbt]
\begin{center}
\caption{Values of the various parameters used in the evolution dynamics.}
\label{initialconditions}
\begin{tabular}{|c|c|}
\hline
$\sqrt{s}$ & 2.76 TeV\\
\hline
$T_{i}$ & 646 MeV\\
\hline
$\tau_{i}$& 0.08 fm\\
\hline
$T_{c}$ & 175 MeV\\
\hline
$T_{ch}$ & 170 MeV\\
\hline
$T_{fo}$ & 120 MeV\\
\hline
EoS & 2+1 Lattice QCD~\cite{MILC}\\
\hline
\end{tabular}

\end{center}
\end{table}

In the present work the space time evolution  is same as done in Ref.~\cite{photon_ans,LB}. 
For $\tau\leq\tau_{iso}$, system evolves anisotropically and is described by 
{\em{free steaming interpolating}} model~\cite{MMprl100,schenke}. The effect of radial flow is 
neglected in the anisotropic phase as it is not developed in the initial stage of the collision. 
For $\tau\geq\tau_{iso}$, the system is described by (1+2)d ideal 
hydrodynamics model with cylindrical symmetry~\cite{hvg} and boost invariance 
along the longitudinal direction~\cite{bjorken}.
The initial temperature ($T_{i}$) and initial formation time ($\tau_{i}$) of the system 
is constrained by the hadronic multiplicity ($dN/dy$) as $dN/dy\sim T_i^3\tau_i$.
The equation of state (EoS) which controls the rate of expansion/cooling
has been taken from the lattice QCD calculations~\cite{MILC}.
The chemical ($T_{ch}$) and kinetic ($T_{fo}$) freeze-out temperatures 
are fixed by the particle ratios and the slope of  $p_T$ 
spectra of hadrons~\cite{hirano}. The values of these parameters are
displayed in Table~\ref{initialconditions}.

\section{Results}
\label{sec_results}

With  the initial conditions described in Table~\ref{initialconditions}, we evaluate the correlation function, 
$C_2$ for different invariant mass windows (for $\langle M \rangle=$0.3, 0.5, 0.77, 1.02, 1.6 and 2.5 GeV) for 
Pb+Pb collisions at $\sqrt{s_{NN}}$ = 2.76 TeV as a function of $q_{side}$ and $q_{out}$ which are related to 
the transverse momentum of individual pair as follows:~\cite{uaw};
\begin{eqnarray}
&&q_{side}=\left|\vec{q_T}-q_{out}\frac{\vec{K_T}}{K_T}\right|\nn\\
&&=\frac{2k_{1T}k_{2T}\sqrt{1-\cos^2(\psi_1-\psi_2)}}
{\sqrt{k_{1T}^2+k_{2T}^2+2k_{1T}k_{2T}\cos(\psi_1-\psi_2)}}\nn\\   
&&q_{out}=\frac{\vec{q_T}.\vec{K_T}}{|K_T|}\nn\\
&&=\frac{(k_{1T}^2-k_{2T}^2)}{\sqrt{k_{1T}^2+k_{2T}^2+2k_{1T}k_{2T}\cos(\psi_1-\psi_2)}}
\label{eq_q}
\end{eqnarray}
where $k_{iT}$ is the individual transverse momentum and $y_i$ is the rapidity. 
It may be mentioned that the BEC function has values 
$1\leq C_{2}(\vec{k_{1}},\vec{k_{2}}) \leq 2$ 
for a chaotic source. These bounds are from
quantum statistics. 

These source dimensions can be obtained
by parameterizing the calculated correlation function with the
empirical Gaussian form~\cite{pratt};
\begin{eqnarray}
&&C_2(q,K)  =1+\lambda \exp(-R^2_iq^2_i) \nn\\
\label{eq_prmt}
\end{eqnarray}
where $i$ stands for side and out. Thus 
$R_{side}$ and $R_{out}$  appearing in Eq.~\ref{eq_prmt},  
are commonly referred to as “HBT radii”, which are measures of 
Gaussian widths of the source size.  
The deviation of $\lambda$ from
1/3 will indicate the presence of non-thermal sources. 
While the radius corresponding to $q_{side}$
($R_{side}$) is closely related to the transverse size of the system. 
The radius corresponding to $q_{out}$ ($R_{out}$) measures both the
transverse size and duration of particle emission ~\cite{uaw,hb3,rischke,hermm,chappm}. 

\begin{figure}[!htbp]
\begin{center}
\includegraphics[scale=0.3]{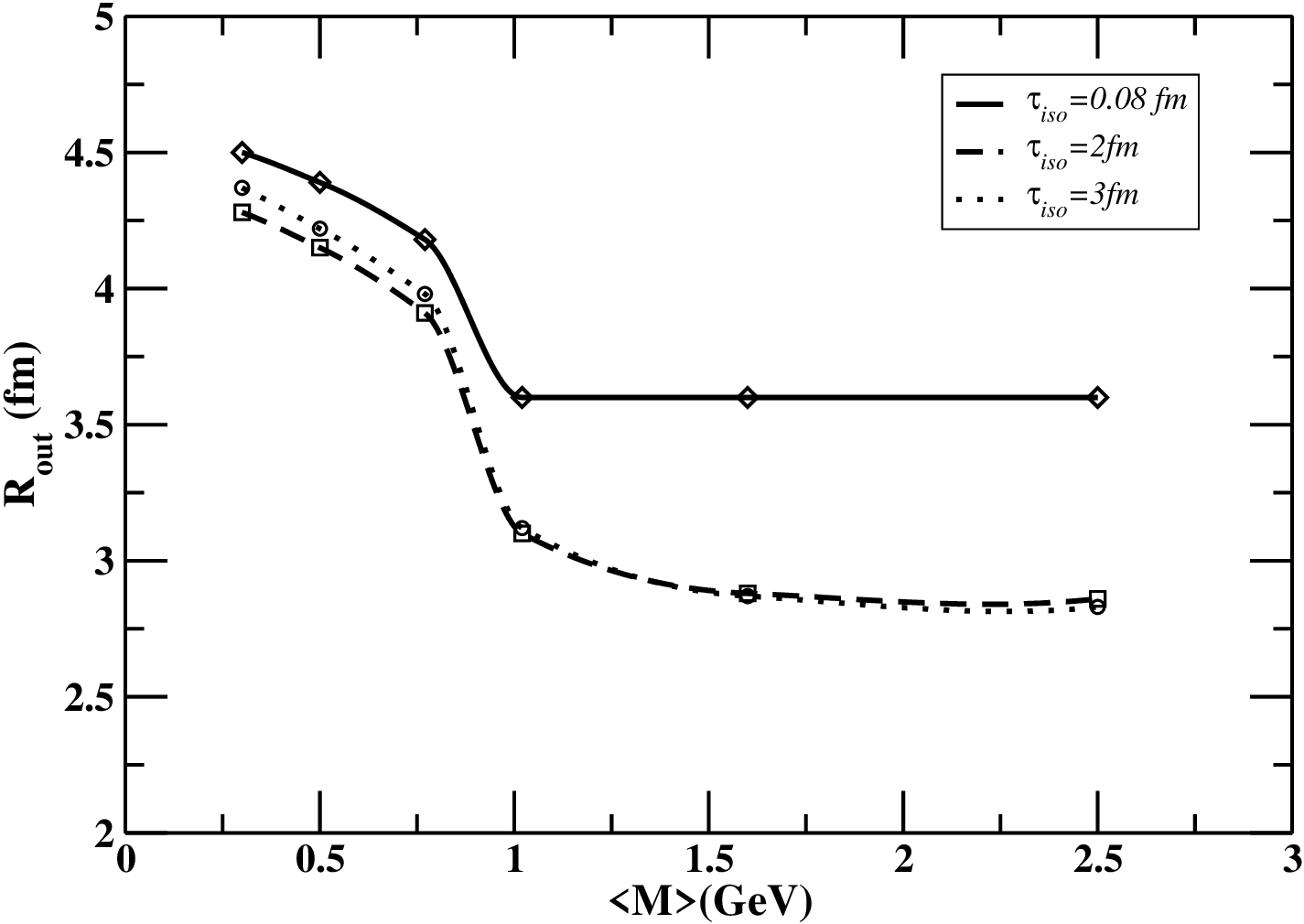}
\end{center}
\caption{The variation of $R_{out}$ with $\langle M\rangle$ for different values of $\tau_{iso}$. 
The solid line corresponds to isotropic scenario ($\tau_{iso}=\tau_i=0.08$ fm) and 
the dotted and dashed line  corresponds to $\tau_{iso}$ = 2,3 fm respectively which 
is related to anisotropic scenario.} 
\label{fig1}
\end{figure}

\begin{figure}[!htbp]
\begin{center}
\includegraphics[scale=0.3]{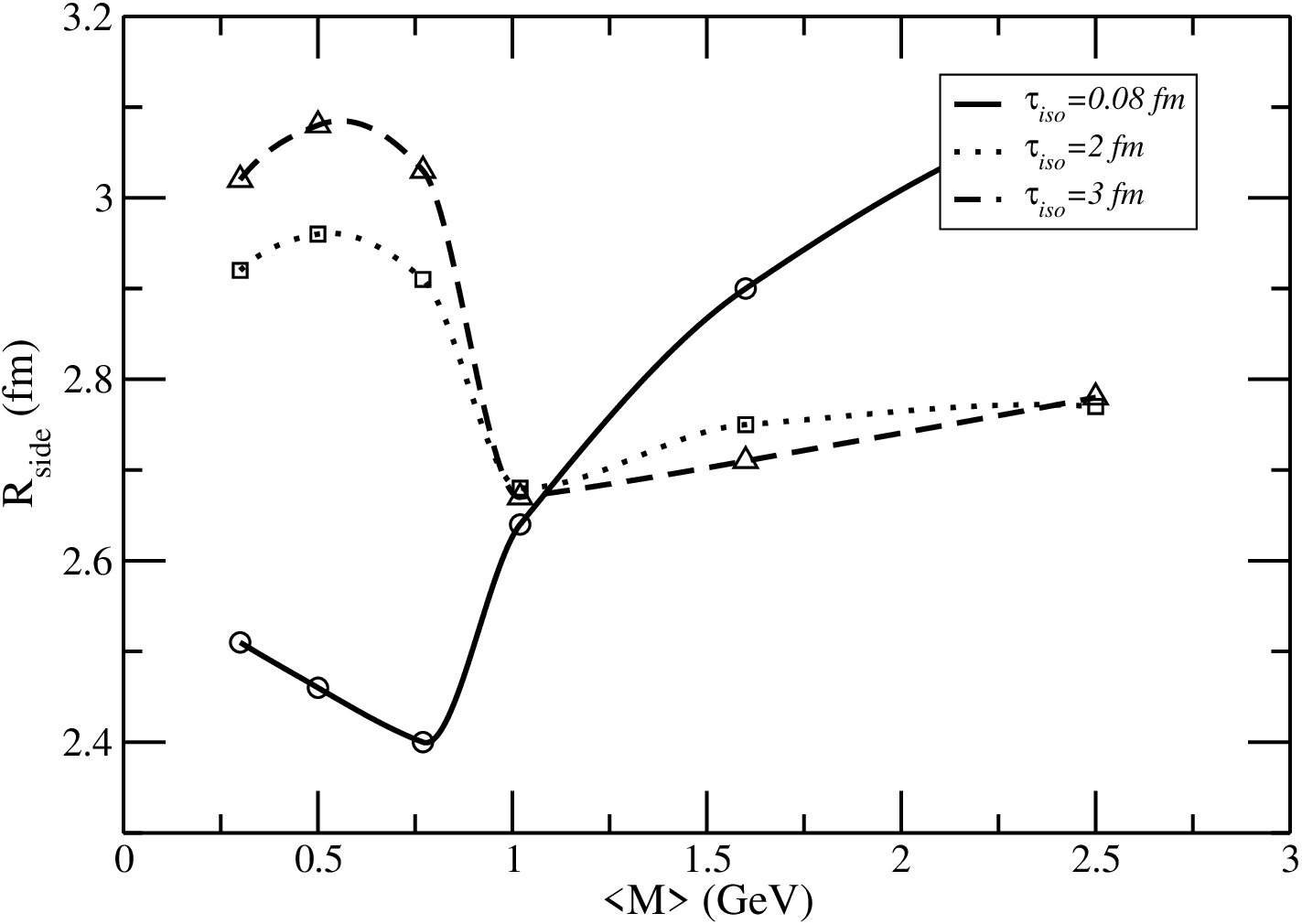}
\end{center}
\caption{The variation of $R_{side}$ with $\langle M\rangle$ for different values of $\tau_{iso}$. 
The solid line corresponds to isotropic scenario ($\tau_{iso}=\tau_i=0.08$ fm) and 
the dotted and dashed line  corresponds to $\tau_{iso}$ = 2,3 fm respectively which 
is related to anisotropic scenario.}  
\label{fig2}
\end{figure}

In the present work, the corresponding  HBT radii are extracted with the help of the parametrization expressed 
in Eq.~\ref{eq_prmt} for three different values of $\tau_{iso}$. We choose $\tau_{iso}$ in such a way 
that  $\tau_{iso}=\tau_{i}$ (0.08 fm) corresponds to the isotropic situation 
and $\tau_{iso}>\tau_{i}$ (2, 3 fm) corresponds to anisotropic scenario. 
So basically, we have attempted to examine the sensitivity of momentum anisotropy on spatial and temporal 
size of the evolving system by controlling the variable, $\tau_{iso}$. 
Again we argue that  large M region (beyond $M\geq m_{\phi}$) corresponds to the partonic phase as it is dominated by 
lepton pairs  from the partonic interactions  and  $M \sim m_\rho$ region corresponds to the hadronic region 
as leptonpairs are produced basically from the interaction of light vector mesons. So the study of  invariant mass 
variations of  $R_{side}$ and $R_{out}$ in these two M regions characterizes the two different phases of HIC~\cite{payal_HBT}. 
In Figure.~\ref{fig1} and ~\ref{fig2} we display the invariant mass dependence of $R_{side}$ and $R_{out}$ 
for three values of $\tau_{iso}=0.08, 2, 3 fm$. 

$C_2$ as function of  $q_{out}$ is calculated by  taking $\psi_1=\psi_2$=0, $y_1=y_2$=0 and fixing 
transverse momentum of one photon ($k_{1T}$ = 2 GeV) and varying the other ($k_{2T}$)  for different invariant  
masses and the values of corresponding $R_{out}$ is extracted  from it using the parametrization given in 
Eq. ~\ref{eq_prmt}.  $R_{out}$ probes both the transverse size and the duration of emission.  
 With the development of radial flow, the transverse dimension of the emission zone decreases.
Although the effect of flow is small in  the initial stage of collision which is dominant in large M region 
(corresponds to larger size) and the duration of emission is small -  
resulting in a small values of $R_{out}$. Whereas  $M \sim m_\rho$ region suffers from larger flow effects which should 
have resulted in a minimum value in $R_{out}$ in this M region. However, $R_{out}$ probes the duration of emission too, 
which is large for hadronic phase because of the slower expansion due to softer EoS used in the present work for the hadronic
phase. The larger duration compensates the reduction of $R_{out}$ due to flow resulting in a bump in 
$R_{out}$ for $M \sim m_{\rho}$ (see Fig. 1). Again by increasing  $\tau_{iso}$, the duration of particle emission shortens 
and results in smaller value of  $R_{out}$.  \\
\\

 $C_2$ as function of  $q_{side}$ is calculated by  taking $k_{1T}$ = $k_{2T}=$ 2 GeV, 
$y_1=y_2$=0 and fixing $\psi_2$=0 and varying $\psi_1$  for different invariant  
masses and the values of corresponding $R_{side}$ is extracted  from it using the parametrization 
given in Eq. ~\ref{eq_prmt}. In Fig ~\ref{fig2} we display the variation of $R_{side}$ with $M$ 
for three different values of $\tau_{iso}$ and observe quantitative as well as qualitative changes in magnitude.   
It can be shown that  $R_{side}$ 
is related to the collective motion of the system through the relation:
$R_{side}\sim 1/(1+E_{\mathrm collective}/E_{\mathrm thermal})$. In large M region, 
as the flow is not developed fully so the values of $R_{side}$ is affected only due to 
initial thermal energy. As $\tau_{iso}$ is inversely related to the temperature, thus 
increasing $\tau_{iso}$ results in decrease in the values of $R_{side}$ in large M region. 
In the mass region corresponds to the hadronic phase, $M \sim m_{\rho}$, the flow is fully developed 
and the thermal energy is reduced. The ratio of collective to thermal energy is large in this M region 
and hence shows smaller $R_{side}$ in isotropic scenario. However, the situation is complex in presence 
of initial momentum space anisotropy. By  increasing $\tau_{iso}$, the flow is reduced resulting in 
larger values of $R_{side}$ in anisotropic scenario.    

 \begin{figure}
\begin{center}
\includegraphics[scale=0.3]{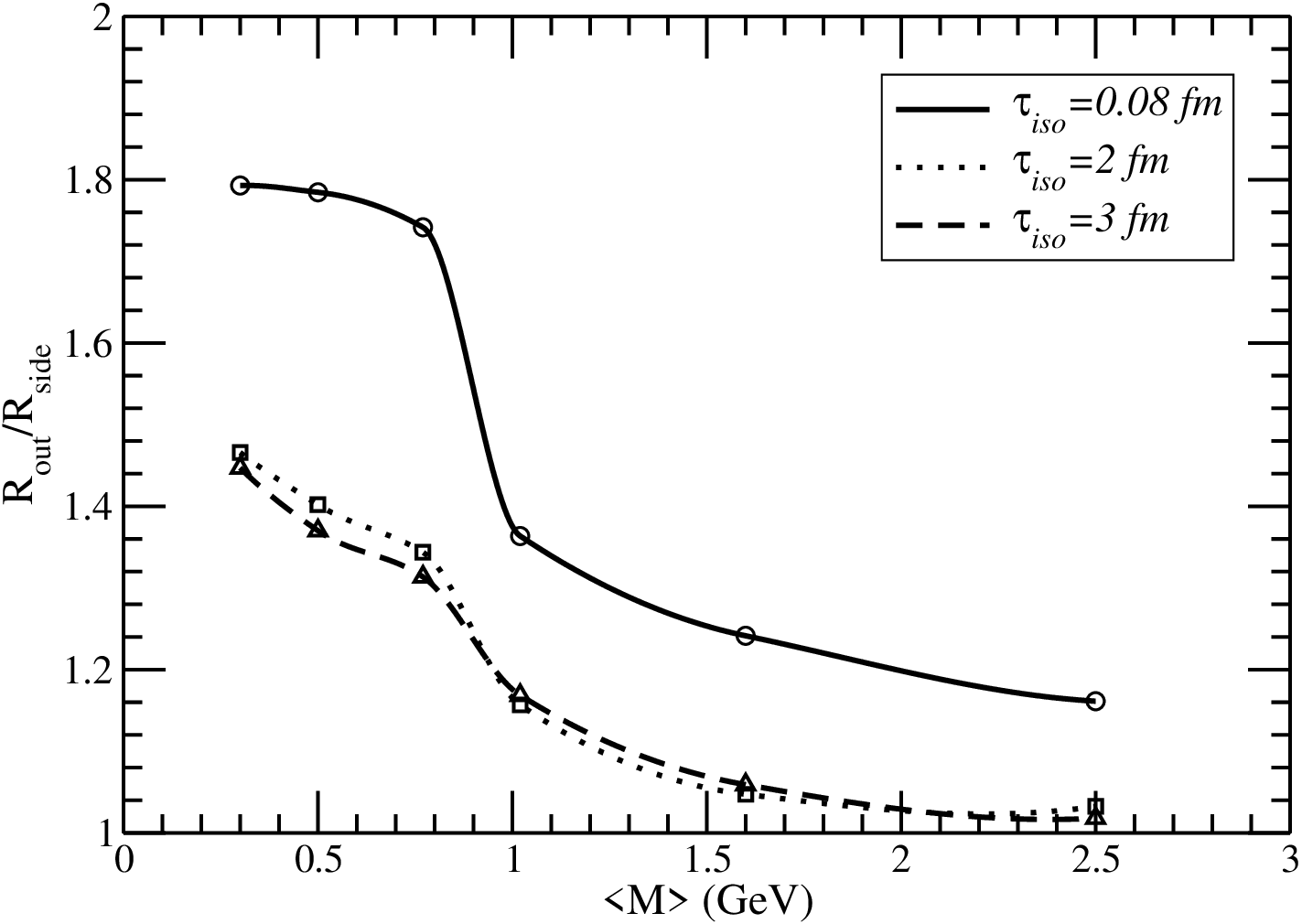}
\end{center}
\caption{The ratio $R_{out}$/$R_{side}$ as a function of  $\langle M \rangle$.}  
\label{fig3}
\end{figure}
The HBT radii are proportional to the average size of the system~\cite{rischke}. 
The average size of the system is related to the HBT radii (here $R_{out}$ and $R_{side}$) 
~\cite{rischke} extracted from correlation function, $C_2$ using Eq.~\ref{eq_prmt}.
However,  some of the model dependence gets canceled out by taking the ratio 
of  $R_{out}$ to $R_{side}$ . Thus the quantity, $R_{\mathrm out}/R_{\mathrm side}$  
gives the duration of particle emission~\cite{pratt,hermm,chappm} 
for various domains of $M$. 

\begin{figure}[!htbp]
\begin{center}
\includegraphics[scale=0.3]{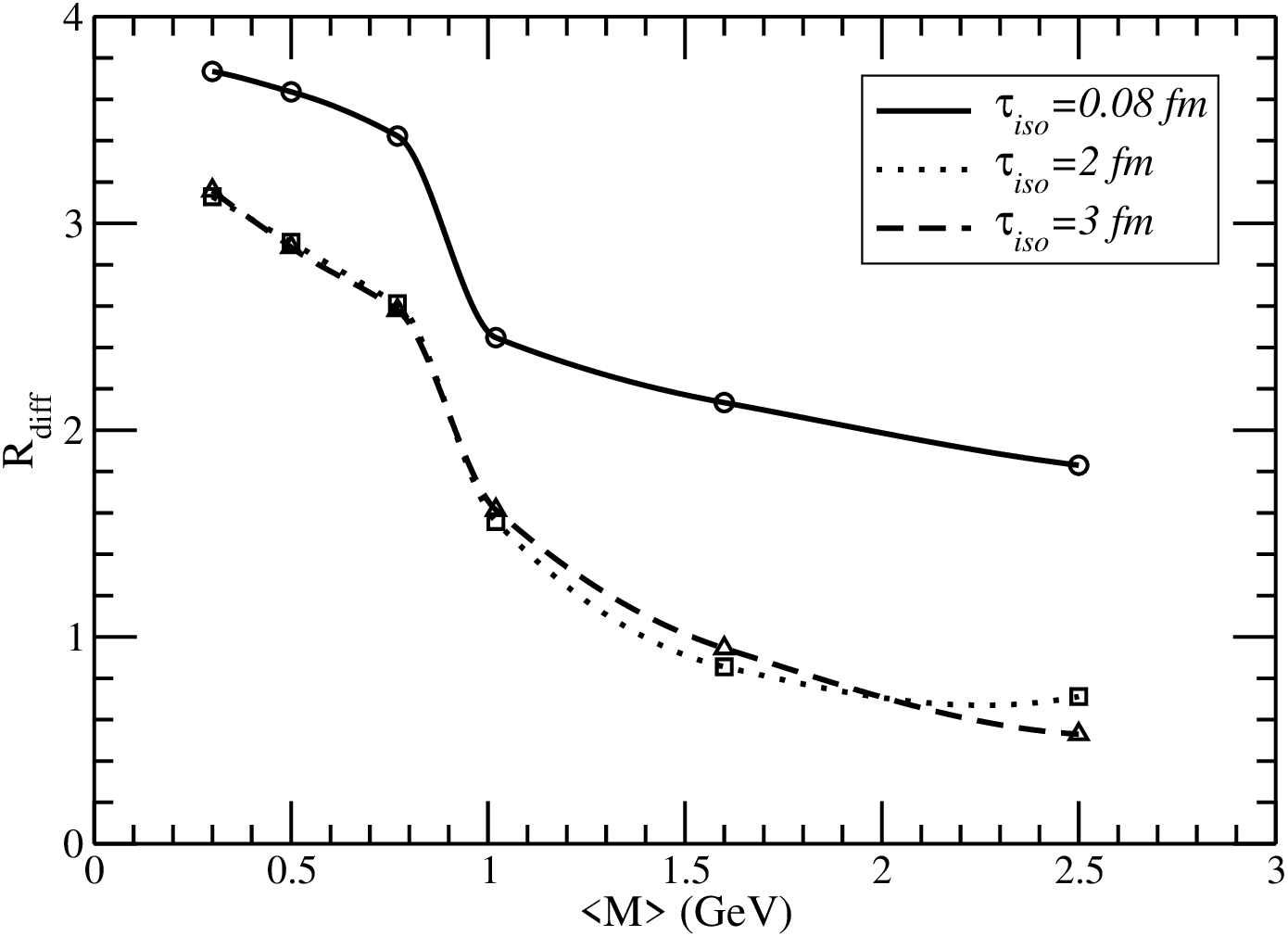}
\end{center}
\caption{The difference $\sqrt{R_{out}^{2} - R_{side}^{2}}$ as a function of  $\langle M \rangle$.}  
\label{fig4}
\end{figure}

 Figures.~\ref{fig3}  and ~\ref{fig4} show 
the variation of  $R_{out}$/$R_{side}$ and $R_{diff}$=
$\sqrt{R_{out}^{2} - R_{side}^{2}}$ as a function of  $\langle M \rangle$  
for $\sqrt{s_{NN}}$ = 2.76 TeV for different values of $\tau_{iso}$. Both show a non-monotonic
dependence on $\langle M \rangle$.  The smaller values of both the quantities,
particularly at high mass region, reflect the contributions from the early partonic phase 
of the system. The peak around $\rho$-meson mass reflects dominance of
the contribution from the late hadronic phase in isotropic scenario. However, increasing 
values of $\tau_{iso}$ result in shorter duration of particle emission hence both 
these quantities have smaller value in anisotropic scenario compared to the isotropic one.
 However, by increasing the values of $\tau_{iso}$  we observe quantitative change in the 
magnitude of both of these quantities. This is because the  duration of particle emission  reduces 
as the system takes more time to become thermalized (by increasing $\tau_{iso}$). Hence both these 
quantities have smaller value in the anisotropic scenario compared to the isotropic one.

\section{Summary}
\label{sec_summ}
In summary, the correlation functions for dilepton pairs have been evaluated for different values of 
$\tau_{iso}$ and the HBT radii have been extracted from it for Pb+Pb collision at 2.76 TeV LHC energies for 
different $\la M\ra$ windows. We observe both qualitative as well as quantitative change  in the variation 
of HBT radii with M for dilepton pairs in presence of the initial momentum anisotropy. We argue that the 
variation of HBT radii with M for dilepton pairs can be used as an efficient tool to follow the change of the 
spatial and temporal dimensions of the evolving system with time  in presence of  momentum space anisotropy 
in the initial stage of collision. The invariant mass dependence of two experimentally challenging quantities, 
$R_{out/R_{side}}$ and $R_{diff}$ also show a quantitative change due to the incorporation of anisotropy in the 
momentum space in the initial stage of collision.  The experimental challenges for the study of dilepton 
interferometry is addressed in ref.~\cite{payal_HBT}, where it has been argued that increasing luminosity of 
the experiment will be a motivating factor for such measurements. 

\section{Acknowledgements}
This work is supported by  Department of Atomic Energy, India. 
PM is partially supported by Korean National Research Foundation.



\begin{thebibliography}{50}
\medskip

\bibitem{pratt} S. Pratt, Phys. Rev. {\bf D 33}, 1314 (1986).

\bibitem{uaw} 
U. A. Weidemann and U. Heinz, Phys. Rep. {\bf 319}, 145 (1999).

\bibitem{hb3} U. Heinz and B. V. Jacak, Ann. Rev. Nucl. Part. Sci. {\bf 49}, 
529 (1999); T. Cs\"orgo and B. L\"orstad, Phys. Rev. C {\bf 54}, 1390 (1996);
B. R. Schlei and N. Xu, Phys. Rev. C {\bf 54}, R2155 (1996); D. H. Rischke and 
M. Gyulassy, Nucl. Phys. A {\bf 608}, 479 (1996).

\bibitem{HBT}
R. Hanbury Brown and R. Q. Twiss,  Nature \textbf{178}, 1046(1956) .

\bibitem{peressounko} D. Peressounko, Phys. Rev. C {\bf 67}, 014905 (2003).

\bibitem{fmann}E. Frodermann, U. Heinz, Phys. Rev. C {\bf 80} 044903 (2009).

\bibitem{AlamHBT}                     
J. Alam, B. Mohanty, P. Roy, S. Sarkar and B. Sinha
 Phys. Rev. C {\bf 67}, 054902 (2003).
 

\bibitem{BMSPRL}
S. A. Bass, B. Mueller and D. K. Srivastava, Phys. Rev. Lett. {\bf 93}, 162301 (2004).

\bibitem{DKS_HBT}
D. K. Srivastava and J. I. Kapusta, Phys. Lett. B \textbf{319}, 407 (1993); 
D. K. Srivastava, Phys. Rev. D {\bf 49}, 4523  (1994).

\bibitem{Rupa_HBT}D. K. Srivastava and R. Chatterjee, Phys. Rev. C {\bf 80}, 054914 (2009).

\bibitem{payal_HBT}
P. Mohanty, J. Alam and B. Mohanty, Phys. Rev. C {\bf 84},  024903 (2011);
P. Mohanty, J. Alam and B. Mohanty, Nucl. Phys. A {\bf 862}  301-303, (2011).
P. Mohanty, J. Alam,  PoS\textbf{(WPCF2011)040}, (2012), arXiV : 1202.2189[Nucl-th].


\bibitem{photon_ans}
P. Mohanty, M. Mandal and P. Roy, Phys. Rev. {\bf C 89}, 054915 (2014).



\bibitem{payal_flow} 
P. Mohanty, J. K. Nayak, J. Alam and S. K. Das, Phys. Rev. C {\bf 82}  034901 (2010); 
J. K . Nayak and J. Alam, Phys. Rev. C 80 (2009) 064906.

\bibitem{payal_v2} 
P.~Mohanty, V.~Roy, S.~Ghosh, S.~K.~Das, B.~Mohanty, S.~Sarkar, J.~e.~Alam and A.~K.~Chaudhuri,
  Phys.\ Rev.\ C {\bf 85} (2012) 031903.
  
\bibitem{MMprl100} M. Martinez and M. Strickland, Phys. Rev. Lett. 100, 102301 (2008). 

\bibitem{mauricio} M. Martinez and M. Strickland, Phys. Rev. {\bf C 78}, 034917 (2008).
 
     
\bibitem{schenke} B. Schenke and M. Strickland Phys. Rev. D {\bf 76}, 025023 (2007).
 
\bibitem{qqpair} J. Cleymans, J. Fingberg and K. Redlich,  Phys. Rev.
D {\bf 35}, 2153 (1987).

\bibitem{emprobe} J. Alam, S. Raha and B. Sinha, Phys. Rep.
{\bf 273}, 243  (1996); J. Alam, S. Sarkar, P. Roy, T. Hatsuda and B. Sinha,
Ann. Phys. {\bf 286}, 159  (2001); R. Rapp and J. Wambach, Adv. Nucl. Phys. {\bf 25}, 1 (2000).


\bibitem{shu} E. V. Shuryak, Rev. Mod. Phys. {\bf 65}, 1  (1993).


\bibitem{4pi} P. Lichard and J. Juran, Phys. Rev. D {\bf 76}, 094030 (2007).

\bibitem{LB} L. Bhattacharya and P. Roy, Phys. Rev. {\bf C 78}, 064904 (2008),
             L. Bhattacharya and P. Roy, Phys. Rev. {\bf C 79}, 054910 (2009).
         

    

\bibitem{MILC} C. Bernard {\it et al.}, Phys. Rev. D {\bf 75}  094505 (2007).

 
             

\bibitem{hvg} H. von Gersdorff, M. Kataja, L. McLerran and P.
V. Ruuskanen, Phys. Rev. D {\bf 34}, 794  (1986). 
             

\bibitem{bjorken} J. D. Bjorken, Phys. Rev. D {\bf 27}, 140 (1983).


\bibitem{hirano} T. Hirano and K. Tsuda, Phys. Rev. C {\bf 66}, 054905 (2002).


  
\bibitem{rischke} D. H. Rischke and M. Gyulassy, Nucl. Phys. A {\bf 608}, 479 (1996).



\bibitem{hermm} 
M. Herrmann and G. F. Bertsch, Phys. Rev. C {\bf 51}, 328 (1995).

\bibitem{chappm} S. Chappman, P. Scotto and U. Heinz, Phys. Rev. Lett. {\bf 74}, 4400 (1995). 

\end{thebibliography}
\end{document}